\documentclass[a4paper,11pt]{article}
\usepackage[utf8]{inputenc}
\usepackage{amsmath}
\usepackage{amssymb}
\usepackage{graphicx}
\usepackage{braket}
\usepackage{geometry}
\usepackage{booktabs}
\usepackage{multirow}
\usepackage{array}
\usepackage{cite}
\usepackage{hyperref}
\usepackage{float}
\usepackage{xcolor}
\usepackage{orcidlink}
\usepackage{lipsum}
\usepackage{tikz}
\usetikzlibrary{shapes,arrows,positioning,decorations.pathmorphing}
\providecommand{\keywords}[1]
{
  \small	
  \textbf{\textit{Keywords---}} #1
}
\title{Entanglement, Coherence, and Recursive Linking in Dicke states:\\
 {\Large A Topological Perspective}}

\author{%
    Sougata Bhattacharyya\textsuperscript{1,2}\thanks{\href{https://orcid.org/0009-0002-9803-4198}{ORCID: 0009-0002-9803-4198}\quad Email: \href{mailto:msc2503121011@iiti.ac.in}{msc2503121011@iiti.ac.in}} \\
    \small Department of Astronomy, Astrophysics and Space Engineering,  \\
    \small Indian Institute of Technology, Indore, 453552, Madhya Pradesh, India
    \and
    Sovik Roy\textsuperscript{2}\thanks{\href{https://orcid.org/0000-0003-4334-341X}{ORCID: 0000-0003-4334-341X}\quad Email: \href{mailto:s.roy2.tmsl@ticollege.org}{s.roy2.tmsl@ticollege.org}} \\
    \small Department of Mathematics, Techno Main Salt Lake, \\
    \small Techno India Group, EM 4/1, Sector V, Salt Lake, Kolkata 700091, India
}
\date{\today}

\begin{document}

\maketitle

\vspace{4cm}
\begin{abstract}
\noindent This work investigates the topological structure of multipartite entanglement in symmetric Dicke states $|D_n^{(k)}\rangle$. By viewing qubits as topological loops, we establish a direct correspondence between the recursive measurement dynamics of Dicke states and the stability of $n$-Hopf links. We utilize the Schmidt rank to quantify bipartite entanglement resilience and introduce the $l_1$-norm of quantum coherence as a measure of \textit{link fluidity}. We demonstrate that unlike fragile states such as $ \left| GHZ \right \rangle$ (analogous to Borromean rings), Dicke states exhibit a robust, self-similar topology where local measurements preserve the global linking structure through non-vanishing residual coherence.
\end{abstract}


\vspace{1cm}

\section{Introduction}

The study of multipartite entanglement remains one of the most challenging and rewarding frontiers in quantum information science \cite{thapliyal1999multipartite, horodecki2009, aczel2002, bengtsson2016brief, ainley2024multipartite}. While bipartite systems are well-understood through the lens of Bell states \cite{bell1964einstein}, systems comprising three or more qubits exhibit a significantly richer structure of correlations that cannot be classified by a single measure \cite{dur2000, amico2008}. Historically, the classification of these states has relied on their transformation properties under Stochastic Local Operations and Classical Communication (SLOCC) \cite{dur2000}. For example, in tripartite systems, the Greenberger-Horne-Zeilinger (GHZ) state \cite{greenberger1990} and the W state \cite{dur2000three} represent two inequivalent classes of genuine tripartite entanglement \cite{dur2000}. However, a parallel and intuitive framework has emerged that seeks to understand these quantum correlations through the lens of topology, specifically, through the mathematical framework of knot theory \cite{kauffman2016, adams1994}. P. K. Aravind pioneered this approach by drawing analogies between the fragility of the GHZ state and the Borromean rings [see Fig.~(1)], where the removal of a single component disentangles the entire system \cite{aravind1997, bhattacharyya2025symmetric}.\\

\noindent In this work, we extend this topological analysis to the family of \textit{Dicke states}, denoted as $|D_n^{(k)}\rangle$. Originally introduced by Robert H. Dicke in 1954 in the context of quantum optics to describe superradiance \cite{dicke1954coherence, bartschi2019deterministic, mukherjee2020preparing}, these states have since found profound relevance in quantum information and multiqubit entanglement studies \cite{li2008, toth2007detection}. Dicke states are a class of multi-qubit (or $n$ qubit) quantum states with a fixed number of excitations, representing a combinatorial superposition of computational basis states. These states are symmetric under the permutation of any two qubits. They are simultaneous eigenstates of the total angular momentum operator $J^2$ and $J_z$, with all qubits treated as spin-$\frac{1}{2}$ particles. Unlike the asymmetric $|Star\rangle$ state \cite{cao2020fragility} or the fragile $|GHZ\rangle$ state, Dicke states exhibit a high degree of permutation symmetry and robustness, a property that has been verified in various experimental realizations using photonic qubits \cite{agrawal2006perfect, kiesel2007experimental, wieczorek2009experimental}. We propose that the entanglement structure of $|D_n^{(k)}\rangle$ maps onto an\textit{ $n$-component Hopf link} (see Fig.~(2), which is three-component Hopf link], where every qubit is pairwise linked to every other qubit. To rigorously establish this analogy, we employ two quantum quantifiers viz. (a) the \textit{Schmidt Rank}, which diagnoses the presence of bipartite entanglement after particle loss (i.e. when measurement is made upon the system) \cite{plenio2007}, and  (b) \textit{Quantum Coherence} (specifically the $l_1$-norm) \cite{cao2020fragility, roy2023exploring}, which we interpret as the \textit{fluidity} that maintains the topological links. We show that the recursive mathematical structure of Dicke states under projective measurement mirrors the recursive preservation of Hopf links, providing a concrete operational bridge between quantum optics and topology. The paper is organized as follows.\\\\
After discussion of theoretical preliminaries required for our study in Sec.~$2$, we show the matematical analysis of the general Dicke state in Sec.~$3$. This is followed by the discussion of measurement dynamics and recursive structure of the Dicke state in Sec.~$4$. Then we discuss about topological analogy in Sec.~$5$ followed by conclusion in Sec.~$6$. Four Appendices $A, B, C$ and $D$ have been added after the reference section.

\section{Theoretical Preliminaries}

In this section, we outline the fundamental concepts necessary for our analysis. We review the definition of Dicke states, the formalism of projective measurements \cite{nielsen2010quantum}, and the two key metrics used to quantify the resulting states viz. Schmidt rank and Quantum Coherence. Finally, we establish the relevant topological definitions \cite{rolfsen1976}.

\subsection{The Dicke State $|D_n^{(k)}\rangle$}
The Dicke state $|D_n^{(k)}\rangle$, an $n-$ qubit quantum state, is defined as the normalized equal superposition of all computational basis states with exactly $k$ excitations (denoted as logical $|1\rangle$s) and $n-k$ ground states (denoted as logical $|0\rangle$s). Mathematically, it is expressed as:
\begin{eqnarray}
    |D_n^{(k)}\rangle = \binom{n}{k}^{-1/2} \sum_{j} P_j \left( |1\rangle^{\otimes k} \otimes |0\rangle^{\otimes (n-k)} \right),
    \label{dicke1}
\end{eqnarray}
where $\sum P_j$ denotes the sum over all distinct permutations of the qubits. As for example, when $n = 3$ and $k = 1$, the Dicke state is $|D_3^{(1)}\rangle$ which is $\frac{|001\rangle + |010\rangle + |100\rangle}{\sqrt{3}}$ i.e. the $|W\rangle$ state \cite{dur2000three}. It is easy to check that swapping any two qubits leaves the state unchanged and hence it is permutation symmetric. Another example of Dicke state is inverted $|W\rangle$ state that corresponds to $|D_3^{(2)}\rangle$ [the inverted $|W\rangle$ state ($|\overline{W}\rangle$) is essentially the bit-flipped version of the $|W\rangle$ state and it is written as $|\overline{W}\rangle = X\otimes X\otimes X|W\rangle$, where $X$ is Pauli-$X$ gate]. The fundamental property of these states is their permutation symmetry while no single qubit plays a privileged role, contrasting sharply with that of graph states like the $|Star\rangle$ state.

\subsection{Projective Measurements}
\noindent A projective measurement on a single qubit is described by the orthogonal projectors $P_0 = |0\rangle\langle 0|$ and $P_1 = |1\rangle\langle 1|$ \cite{nielsen2010quantum}. For a multipartite state $|\psi\rangle$, the probability of obtaining outcome $m \in \{0, 1\}$ on a specific qubit is given by the Born rule:
\begin{eqnarray}
    p(m) = \langle \psi | P_m | \psi \rangle.
    \label{prob1}
\end{eqnarray}

\noindent The post-measurement state of the remaining system, $|\psi^{(m)}\rangle$, is obtained by projecting and renormalizing and is given as
\begin{eqnarray}
    |\psi^{(m)}\rangle = \frac{P_m |\psi\rangle}{\sqrt{p(m)}}.
    \label{state1}
\end{eqnarray}

\noindent In our topological analysis, this measurement process is the quantum analogue of \textit{cutting} \cite{mukherjee2020preparing} or \textit{removing a ring from a link}. The properties of $|\psi^{(m)}\rangle$ determine the stability of the remaining topological structure \cite{radhakrishnan2024entanglement}.

\subsection{Schmidt Rank as an Entanglement Measure}
To determine if the remaining qubits stay linked after a measurement, we utilize the Schmidt decomposition \cite{peres1995, nielsen2010quantum}. A pure bipartite state $|\phi\rangle_{AB}$ can be written as:
\begin{eqnarray}
    |\phi\rangle_{AB} = \sum_{i=1}^{R} \lambda_i |u_i\rangle_A \otimes |v_i\rangle_B,
    \label{schmidt1}
\end{eqnarray}
where $\lambda_i > 0$ are the Schmidt coefficients satisfying $\sum \lambda_i^2 = 1$. The integer $R$ is the \textit{Schmidt Rank}.
\begin{itemize}
    \item If $R=1$, the state is \textit{separable} (product state).
    \item If $R > 1$, the state is \textit{entangled} \cite{horodecki2009}.
    \item If $R=2$ with $\lambda_1 = \lambda_2 $, it indicates a \textit{maximally entangled} bipartite state (such as Bell states).
\end{itemize}

\subsection{Quantum Coherence}
While Schmidt rank indicates the \textit{existence} of entanglement, quantum coherence shows the \textit{strength} of the superposition that supports it. It is to be noted that, Quantum Coherence is basis-dependent, i.e., when coherence is measured in different bases, the quantified value is different \cite{roy2023exploring}. In this work, we have quantified coherence in the computational basis. To quantify coherence, we employ the $l_1$-norm measure of coherence. While multiple coherence measures exist, the $l_1$-norm offers both theoretical relevance and computational tractability. Now, for a density matrix $\rho$, the measure $l_1-$ norm is defined as the sum of the absolute values of the off-diagonal elements \cite{roy2023exploring}. This measure was rigorously quantified within the resource theoretic framework by Baumgratz et al. \cite{baumgratz2014quantifying}. Denoting $l_1-$ norm of coherence by $C_{l_1}(\rho)$, for a density operator $\rho$ with respect to a given state the said norm is defined as,
\begin{eqnarray}
    C_{l_1}(\rho) = \sum_{i \neq j} |\rho_{ij}|.
    \label{coherence1}
\end{eqnarray}
For a pure state $|\psi\rangle = \sum c_i |i\rangle$, the Eq.(\ref{coherence1}) simplifies to the following form (see Appendix \ref{app:coherence_derivation} for the detailed derivation):
\begin{eqnarray}
    C_{l_1}(|\psi\rangle) = \left( \sum_i |c_i| \right)^2 - 1.
    \label{coherence2}
\end{eqnarray} 

\subsection{Topological Links}
We utilize standard definitions from knot theory \cite{adams1994, rolfsen1976} to classify the quantum states:
\begin{itemize}
    \item \textbf{Unlink:} A collection of loops that are not topologically connected (analogous to a product state, i.e. the state with Schmidt rank $R=1$).
    \begin{figure}[H]
\centering
\label{boro}
\includegraphics[scale=0.3]{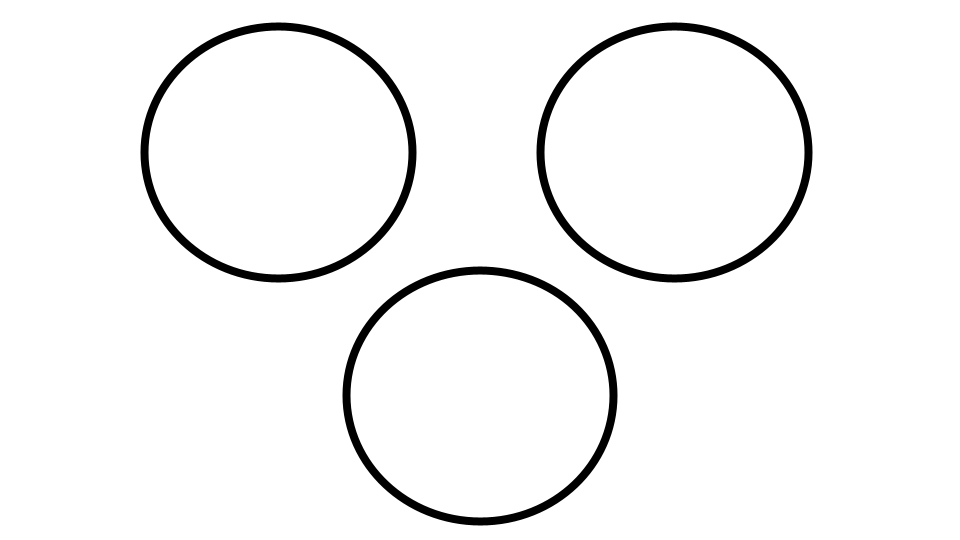}
\caption{Unlink}
\end{figure}

    \item \textbf{Borromean Rings:} A set of three rings where no two are linked, but the three together are inseparable. Cutting one releases the other two (analogous to the GHZ state \cite{aravind1997,mukherjee2020preparing}).

        \begin{figure}[H]
\centering
\label{boro}
\includegraphics[scale=0.3]{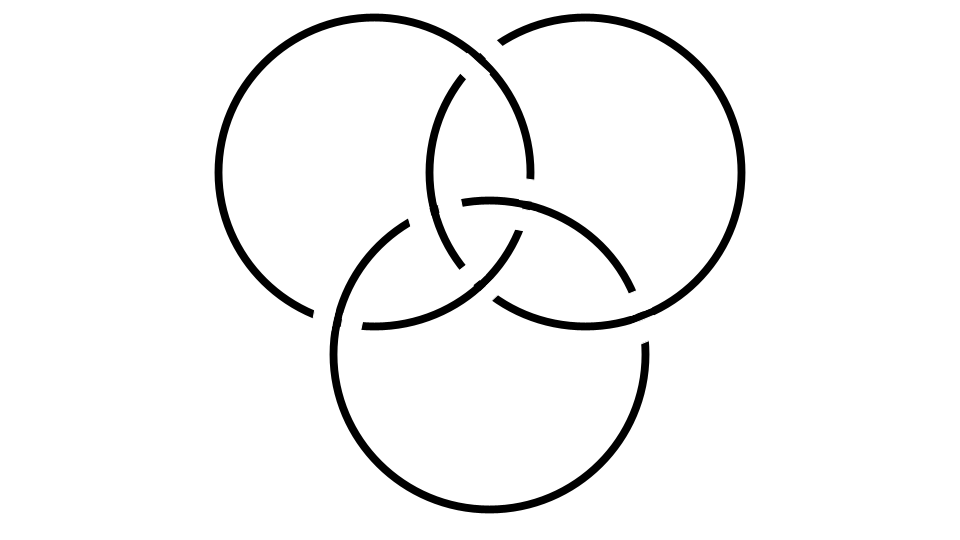}
\caption{Borromean Rings}
\end{figure}

    \item \textbf{Hopf Link:} The simplest non-trivial link between two rings.

    \begin{figure}[H]
\centering
\label{boro}
\includegraphics[scale=0.25]{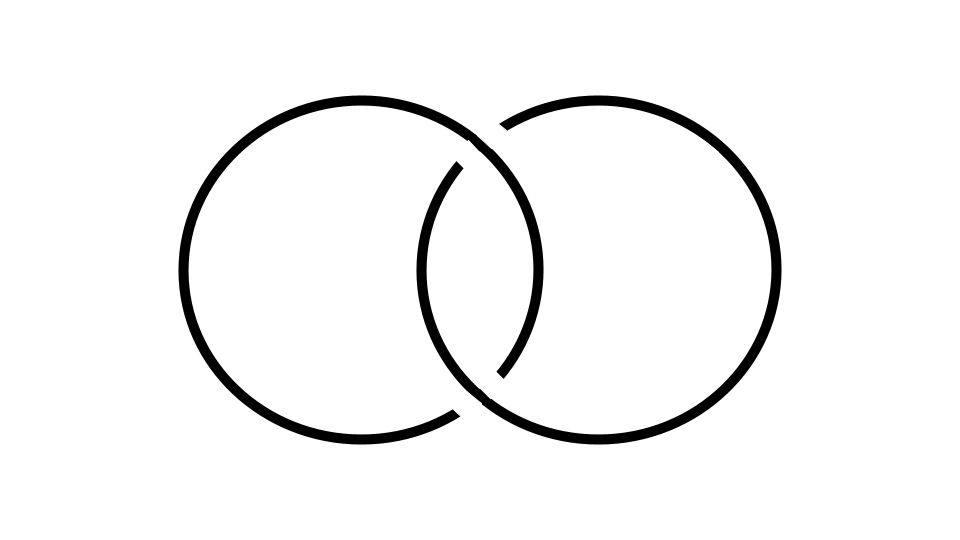}
\caption{Hopf Link}
\end{figure}

    \item \textbf{$n$-Hopf Link:} A generalization where every component is pairwise linked to every other component. Cutting one component leaves a robust $(n-1)$-component structure intact. We posit this as the topological equivalent of the Dicke state \cite{li2008,mukherjee2020preparing}.
    \begin{figure}[H]
\centering
\label{boro}
\includegraphics[scale=0.3]{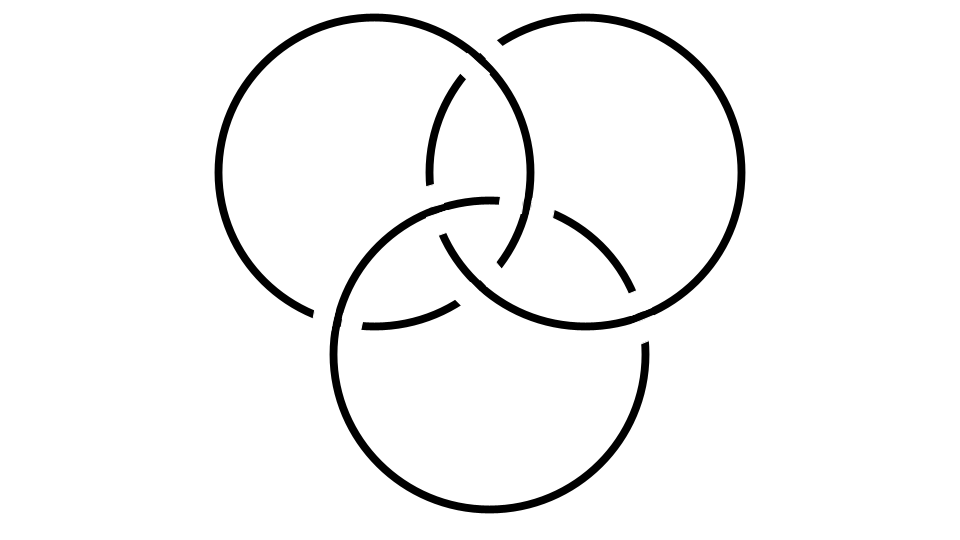}
\caption{3-Hopf Link}
\end{figure}
\end{itemize}

\subsection{Link Fluidity: The Mechanism of Topological Preservation}

Although the Schmidt rank provides a binary diagnostic for the presence of entanglement (i.e., linked vs. unlinked), it does not explain the \textit{mechanism} that preserves this structure under the destructive operation of measurement. To describe this mechanism, we introduce the concept of \textit{Link Fluidity} or \textit{Fluidity}.

\subsubsection{Link Fluidity via Quantum Coherence}
In classical knot theory, a physical link is often viewed as a static, localized connection between components \cite{adams1994}. However, in the quantum regime of Dicke states, the \textit{link} is not a localized bond but a \textit{delocalized correlation} distributed across the entire Hilbert space. We define \textit{Link Fluidity} ($\mathcal{F}$) as the capacity of the entangled structure to redistribute its correlations dynamically upon the loss of a subsystem. Mathematically, this is quantified by the $l_1$-norm of quantum coherence, expressed in the computational basis as
\begin{eqnarray}
    \mathcal{F} \equiv C_{l_1}(|D_n^{(k)}\rangle) = \binom{n}{k} - 1.
    \label{coherence3}
\end{eqnarray}

\subsubsection{Conceptual Framework: Rigid vs. Fluid Entanglement}

To facilitate our topological analysis, we introduce a theoretical distinction between two regimes of multipartite entanglement, governed by the magnitude of their quantum coherence. We propose that the stability of a topological link under measurement is determined by the \textit{fluidity} of its support in the Hilbert space \cite{cao2020fragility}. From Eq.(\ref{coherence3}), it is clear that the link fluidity or fluidity of a multipartite quantum state is directly proportional to coherence of the state.

\begin{itemize}
    \item \textbf{Rigid Entanglement in the low coherence regime:} 
    We characterize states with very little quantum overlap as possessing \textit{rigid entanglement}. In this scenario, the connection between particles relies on just a few specific links. It is like a bridge supported by only a single pillar. It is a brittle structure where everything depends on that one support holding up. Because there are no extra supports or backups, the structure cannot handle any local \textit{damage}. Consequently, if you remove just one part (via measurement), the entire connection breaks, and the system instantly falls apart \cite{radhakrishnan2024entanglement}.

    \item \textbf{Fluid Entanglement in the high coherence regime:} 
    On the other hand, we characterize states with high quantum overlap as possessing \textit{fluid entanglement}. In this regime, the connection is spread out across a vast network of many different links. It behaves less like a stiff chain and more like a spiderweb or a fluid mesh. This setup has high \textit{redundancy}, meaning there are thousands of alternative pathways connecting the system. We believe this flexibility allows the connection to survive if one part is lost. Just as water flows to fill a gap, the remaining links redistribute the load, keeping the system connected even after a part is removed \cite{roy2025environment}.
\end{itemize}

\subsection{The Operational-Topological Framework}

To analyze the structural stability of Dicke states of Eq.(\ref{dicke1}), we establish a unified framework connecting quantum operations to topological concepts. We propose that the following four elements offer complementary perspectives on the same underlying physical reality:

\begin{enumerate}
    \item \textbf{Linking (The Geometry):} 
This provides the qualitative model. It describes the global connectivity pattern of the qubits (e.g., $n$-Hopf link vs. Borromean rings) and defines the system's structural class \cite{kauffman2016}.

\item\textbf{Projective Measurement (The Surgery):} This serves as the \textit{operational probe} \cite{nielsen2010quantum}. In this framework, the quantum collapse of a single qubit maps directly to the topological deletion of a single loop. Thus we have
\begin{eqnarray}
         \mathcal{M}_{\text{proj}} \iff \text{Topological Component Deletion (Cutting)}.
         \label{proj1}
\end{eqnarray}
This operation acts as a \textit{stress test}, revealing whether the entanglement structure is \textit{robust} (recursive) or \textit{fragile} (catastrophic).

\item \textbf{Schmidt Rank (The Indicator):} This serves as the \textit{existence condition} for the residual link after the surgery is performed \cite{plenio2007}.
\begin{equation}
    R = 1 \iff \text{Link Broken (Fragile/Borromean)}
\end{equation}
\begin{equation}
    R > 1 \iff \text{Link Preserved (Robust/Hopf)}
\end{equation}
    
\item\textbf{Quantum Coherence (The Stability Measure):} This serves as the \textit{mechanism} of \textit{Link Fluidity}. It quantifies the delocalization of the link across the Hilbert space, explaining \textit{why} the connection survives the surgery \cite{roy2023exploring}.
\begin{eqnarray}
    C_{l_1} \gg 0 \implies \text{Fluid Entanglement (Recursive Preservation)}
    \label{coherence4}
\end{eqnarray}
High coherence implies that the \textit{links} between qubits are not rigid but fluidly distributed across the system, contributing to topological robustness.

\end{enumerate}
\section{Mathematical Analysis of the General Dicke State $|D_n^{(k)}\rangle$}

Now, we analyze the properties of the general Dicke state before any measurement is performed. We derive its specific algebraic form and calculate its initial quantum coherence, establishing a baseline for the topological analysis.

\subsection{Initial State Properties}
We consider a general Dicke state describing a system of $n$ qubits sharing $k$ excitations. As defined in the literature \cite{agrawal2006perfect,quinta2019cut,bartschi2019deterministic} and our preliminary notes, the state $|D_n^{(k)}\rangle$ is the symmetric superposition of all distinct permutations of $k$ excitations ($|1\rangle$) and $n-k$ ground states ($|0\rangle$).

Mathematically, the state is expressed as:
\begin{eqnarray}
    |D_n^{(k)}\rangle = \frac{1}{\sqrt{\binom{n}{k}}} \sum_{j} P_j \left( |1\rangle^{\otimes k} \otimes |0\rangle^{\otimes (n-k)} \right).
    \label{dicke2}
\end{eqnarray}
where the summation runs over all distinct permutations $P_j$ of the qubits. We defined this state earlier in Eq.(\ref{dicke1}), yet we would like to write it once more in this section.
\begin{itemize}
    \item The total number of terms in the superposition corresponds to the binomial coefficient $\binom{n}{k}$.
    \item The normalization factor is $\mathcal{N} = \frac{1}{\sqrt{\binom{n}{k}}}$.
    \item In the computational basis, every basis state appearing in the superposition has an identical probability amplitude of $c_i = \frac{1}{\sqrt{\binom{n}{k}}}$.
\end{itemize}

\subsection{Calculation of Initial Quantum Coherence}
We now quantify the \textit{link fluidity} of this topological structure by calculating the $l_1$-norm of coherence. Using the formula given in Eq.(\ref{coherence2}) \footnote{derived in Appendix A for pure states} we proceed as follows:\\

\textit{Basis Coefficients:}
	The state contains $\binom{n}{k}$ computational basis states. For each such state $|i\rangle$, the coefficient is:
    \begin{eqnarray}
        c_i = \frac{1}{\sqrt{\binom{n}{k}}}.
        \label{binom1}
    \end{eqnarray}
    
\textit{Sum of Absolute Coefficients:}
	Summing over all $\binom{n}{k}$ terms:
    \begin{eqnarray}
        \sum_{i} |c_i| = \binom{n}{k} \times \left( \frac{1}{\sqrt{\binom{n}{k}}} \right) = \sqrt{\binom{n}{k}}.
        \label{binom2}
    \end{eqnarray}
    
\textit{Coherence:}
	Substituting this sum into the coherence formula of Eq.(\ref{coherence2}):
    \begin{eqnarray}
        C_{l_1}(|D_n^{(k)}\rangle) = \left( \sqrt{\binom{n}{k}} \right)^2 - 1 = \binom{n}{k} - 1.
        \label{binom3}
    \end{eqnarray}

\textit{Result:}
\begin{eqnarray}
    C_{l_1}(|D_n^{(k)}\rangle) = \binom{n}{k} - 1.
    \label{binom3}
\end{eqnarray}

\noindent Topologically, this magnitude of Eq.(\ref{binom3}) represents the \textit{total capacity of the $n$-Hopf link} to distribute stress (or loss) across its components. The maximum coherence occurs at $k \approx n/2$, corresponding to the most robustly linked configuration.

\section{Measurement Dynamics and Recursive Structure of the Dicke State $|D_n^{(k)}\rangle$:}

We now present the analytical results regarding the entanglement and coherence properties of the Dicke state $|D_n^{(k)}\rangle$ under local projective measurement. The measurement is performed on a single qubit (designated as qubit 1) in the computational basis $\{|0\rangle, |1\rangle\}$.

\subsection{Measurement Results and State Evolution:}
The projective measurement yields two possible outcomes, $|0\rangle$ or $|1\rangle$. The post-measurement states of the remaining $n-1$ qubits preserve the Dicke structure but with modified parameters \cite{radhakrishnan2024entanglement}. The complete analytical profile, including the Schmidt rank of the bipartite cut (qubit 1 vs. rest) and the $l_1$-norm of quantum coherence, is summarized in Table \ref{tab:dicke_comprehensive}.

\begin{table}[h]
\centering
\renewcommand{\arraystretch}{1.5}
\resizebox{\textwidth}{!}{%
\begin{tabular}{|c|c|l|c|c|c|c|}
\hline
\textbf{Case} & \textbf{Regime} & \textbf{Outcome} & \textbf{Prob.} & \textbf{Residual State} & \textbf{Schmidt Rank} & \textbf{Coherence ($C_{l_1}$)} \\
\hline
\textbf{Initial State} & All $k$ & -- & -- & $|D_n^{(k)}\rangle$ & -- & $\binom{n}{k} - 1$ \\
\hline
\multirow{2}{*}{\textbf{Entangled}} & \multirow{2}{*}{$0 < k < n$} & $|0\rangle$ & $\frac{n-k}{n}$ & $|D_{n-1}^{(k)}\rangle$ & 2 & $\binom{n-1}{k} - 1$ \\
\cline{3-7}
& & $|1\rangle$ & $\frac{k}{n}$ & $|D_{n-1}^{(k-1)}\rangle$ & 2 & $\binom{n-1}{k-1} - 1$ \\
\hline
\multirow{2}{*}{\textbf{Separable}} & $k=0$ & $|0\rangle$ & 1 & $|0\rangle^{\otimes (n-1)}$ & 1 & 0 \\
\cline{2-7}
& $k=n$ & $|1\rangle$ & 1 & $|1\rangle^{\otimes (n-1)}$ & 1 & 0 \\
\hline
\end{tabular}%
}
\caption{Measurement outcomes for single-qubit projective measurement on Dicke state $|D_n^{(k)}\rangle$. For entangled states, the residual state depends on the measurement outcome, either maintaining $k$ excitations or reducing to $k-1$.}
\label{tab:dicke_comprehensive}
\end{table}

\subsection{Schmidt Rank Analysis}
We analyze the entanglement across the bipartite cut between qubit $1$ and the remaining $n-1$ qubits.

\subsubsection{Boundary Cases ($k=0$ and $k=n$)}
\begin{itemize}
    \item If $k=0$, the state is $|0\rangle^{\otimes n}$. This is a product state whose Schmidt Rank $R = 1 $.
    \item If $k=n$, the state is $|1\rangle^{\otimes n}$. This is also a product state whose Schmidt Rank $ R = 1 $.
\end{itemize}

\subsubsection{General Case ($0 < k < n$)}
For any intermediate number of excitations, the state can be decomposed as a superposition of two orthogonal branches:
\begin{eqnarray}
    |D_n^{(k)}\rangle = \sqrt{\frac{n-k}{n}} |0_1\rangle \otimes |D_{n-1}^{(k)}\rangle + \sqrt{\frac{k}{n}} |1_1\rangle \otimes |D_{n-1}^{(k-1)}\rangle.
    \label{dicke3}
\end{eqnarray}
Since the states $|D_{n-1}^{(k)}\rangle$ and $|D_{n-1}^{(k-1)}\rangle$ are orthogonal (their \textit{total excitation counts} are different), this form constitutes a valid Schmidt decomposition \cite{peres1995}.
\begin{itemize}
    \item There are exactly two non-zero Schmidt coefficients: $\lambda_1 = \sqrt{\frac{n-k}{n}}$ and $\lambda_2 = \sqrt{\frac{k}{n}}$.
    \item Therefore, the Schmidt Rank is strictly $R=2$.
    \item This confirms that for any $0 < k < n$, the measured qubit is entangled with the rest of the system.
\end{itemize}

\subsection{Quantum Coherence Analysis}
The Quantum Coherence ($l_1$-norm) measures the superposition strength of the state in the computational basis.
\begin{itemize}
    \item \textbf{Initial Coherence:} Before measurement, the coherence is $C_{l_1} = \binom{n}{k} - 1$.
    \item \textbf{Residual Coherence (when measurement outcome is $|0\rangle$):} The remaining state is $|D_{n-1}^{(k)}\rangle$. The coherence becomes $\binom{n-1}{k} - 1$.
    \item \textbf{Residual Coherence (when measurement outcome is $|1\rangle$):} The remaining state is $|D_{n-1}^{(k-1)}\rangle$. The coherence becomes $\binom{n-1}{k-1} - 1$.\\
\end{itemize}
Crucially, in the entangled regime ($0 < k < n$), both residual coherence values are strictly positive, indicating that the remaining system retains quantum superposition \cite{roy2025environment}.

\section{Topological Analogy: The Recursive $n$-Hopf Link}

The defining feature of the Dicke state $|D_n^{(k)}\rangle$ is its complete permutation symmetry where every qubit contributes identically to the global entanglement structure. Translating this into topology, we view each qubit as a topological loop (or ring). Unlike asymmetric states such as the Star configuration \cite{cao2020fragility}, where specific qubits play central or peripheral roles, the Dicke state corresponds to a system of $n$ identical loops, all mutually interlinked with equal strength. We identify this structure as an \textit{$n$-Hopf link}. In a topological sense, the entire system can be pictured as $n$ loops where there is no privileged connection between any specific pair; instead, the \textit{linking} is uniform across the manifold \cite{kauffman2016}.

\subsection{Excitation Number $k$ as Link Density}
The integer $k$, representing the number of excitations in the Dicke state, acts as a control parameter for how entanglement is distributed across the qubits. Topologically, we interpret $k$ as the \textit{Linking Density} among the loops.

\subsubsection{The Separable Regime ($k=0$ and $k=n$): Zero Link Density}
At the extreme values, the state occupies $|0\rangle^{\otimes n}$ or $|1\rangle^{\otimes n}$. In these configurations, there are no superpositions and no shared excitations.
\begin{itemize}
    \item \textbf{Topological Interpretation:} This corresponds to $n$ completely disjoint and unlinked loops. Each loop exists independently, with no crossings or linkages between them \cite{adams1994}.
    \item \textbf{Quantum State Properties:} The Schmidt Rank is 1, and the Quantum Coherence is zero ($C_{l_1} = 0$). Removing or manipulating one loop has no effect on the others, representing the \textit{zero link density} configuration.
\end{itemize}

\subsubsection{The Linked Regime ($0 < k < n$): The $n$-Hopf Link Structure}
For any non-extreme excitation number, the state becomes a superposition where excitations are delocalized across different qubits. This shared delocalization gives rise to non-trivial quantum correlations among all subsystems.
\begin{itemize}
    \item \textbf{Topological Interpretation:} The loops are now interlinked. The resulting structure is an $n$-Hopf-Linked configuration, where every pair is symmetrically linked \cite{li2008}.
    \item \textbf{Link Fluidity:} We have already introduced the concept of link fluidity in sec.~$2.6.1$  which will be needed to describe this connection. Unlike a static chain, the linking is mediated by the \textit{quantum coherence} ($C_{l_1} > 0$). The links are \textit{fluidly} distributed across the system, allowing the structure to absorb the loss of a component without total collapse.
\end{itemize}

\subsubsection{Maximally Dense Linking ($k \approx \frac{n}{2}$)}
As derived in Section 4, the state achieves maximal bipartite entanglement when $k = n/2$.
\begin{itemize}
    \item \textbf{Topological Interpretation:} This corresponds to the point in the family of links where the connectivity is maximally dense. The connection between any single ring and the collective of the others is as strong as possible.
    \item \textbf{Resilience:} The entire structure remains highly resilient to the loss of any single component, as the remaining rings are supported by the maximum possible combinatorial coherence $\binom{n}{\frac{n}{2}}-1$.
\end{itemize}

\noindent
Thus, the excitation number $k$ serves as a tunable parameter for the topological density of the system, illustrated in Table \ref{tab:topology_class}.\\\\

\begin{table}[h]
\centering
\renewcommand{\arraystretch}{1.5}
\begin{tabular}{|c|c|c|c|}
\hline
\textbf{Excitation $k$} & \textbf{Quantum State} & \textbf{Topological Analogue} & \textbf{Connectivity} \\
\hline
$k=0, n$ & Product State & Unlinked Loops & Disconnected \\
\hline
$k=1$ & W State & 1-Dense Hopf Link & Sparse (Fragile Coherence) \\
\hline
$k=\frac{n}{2}$ & Balanced Dicke & Maximally Dense Hopf Link & Robust (Maximal Coherence) \\
\hline
\end{tabular}
\caption{Topological classification of Dicke states based on excitation number. The case $k = \frac{n}{2}$ represents the most tightly woven topological structure, corresponding to maximal bipartite entanglement.}
\label{tab:topology_class}
\end{table}

\noindent In summary, the Dicke state $|D_n^{(k)}\rangle$ is the quantum embodiment of a robust, fluidly linked $n$-Hopf bundle. Its ability to maintain Schmidt rank 2 and non-zero coherence through recursive measurements confirms that its topology is protected not by specific pairwise bonds, but by the symmetric distribution of entanglement across the entire manifold of qubits.

\subsection{Quantum coherence as link fluidity}
While the Schmidt rank indicates that the links persist, the quantum coherence ($C_{l_1}$) explains how they persist. We have already introduced the concept of Link Fluidity in the sec.~$2.6.1$ and it have been expressed in the Eq.(\ref{binom3}) as $C_{l_1}(|D_n^{(k)}\rangle) = \binom{n}{k} - 1$. In a static topological link (like a steel chain), the connection is localized \cite{rolfsen1976}. However, in the quantum Dicke state, the \textit{linking} is mediated by the delocalization of the $k$ excitations across the $n$ qubits. The coherence measure $C_{l_1}$ quantifies this delocalization.\\\\
\noindent We interpret the binomial coefficient $\binom{n}{k}$ as the \textit{Topological Volume} of the state which is \textit{the number of distinct pathways through which the qubits are connected.}
\begin{itemize}
    \item \textbf{High Coherence ($k \approx \frac{n}{2}$):} In this situation, the fluidity is maximal. The links are distributed across a vast combinatorial space, making the global entanglement highly robust against the loss of a single site.
    \item \textbf{Residual Coherence:} Crucially, our derivation shows that after measurement, the residual coherence $C_{l_1}' = \binom{n-1}{k'} - 1$ remains strictly positive (for $0 < k' < n-1$). This non-vanishing coherence acts as the \textit{topological glue}, preventing the remaining loops from drifting apart into a product state.
\end{itemize}

\subsection{Topological Stability and Recursive Preservation}
A remarkable property of Dicke states is their stability under local measurements. Topologically, a projective measurement corresponds to \textit{cutting} or \textit{removing} one loop from the link configuration. For the $|GHZ\rangle$ state (analogous to Borromean rings \cite{aravind1997}), cutting one ring destroys the entire structure. However, for the Dicke state, this operation does not destroy the overall linkage. Instead, it results in a smaller but topologically similar connectivity pattern. This recursive preservation gives rise to a self-similar hierarchy:
\begin{eqnarray}
\label{pattern}
    \text{n-loop Dicke Link} \xrightarrow{\text{Cut}} \text{(n-1)-loop Dicke Link} \xrightarrow{\text{Cut}} \text{(n-2)-loop Dicke Link} \xrightarrow{\text{Cut}} \dots
\end{eqnarray}

\noindent Dicke states thus realize a \textit{self-replicating topological family}. This is analogous to a recursive link construction in knot theory, where successive removal of components still leaves smaller, intact links. This behavior is mathematically guaranteed by the non-vanishing residual coherence derived in our analysis, which ensures the topological glue persists even after surgery.

\section{Conclusion}

In this work, we have presented a comprehensive topological classification of symmetric Dicke states $|D_{n}^{(k)}\rangle$, establishing them as the quantum mechanical analogue of the $n-$ component Hopf link. By integrating concepts from knot theory with quantum information measures, we have provided an operational framework that distinguishes the robust connectivity of Dicke states from the fragile Borromean entanglement characterizing GHZ states. Our analysis relied on two complementary quantifiers: the Schmidt rank and the $l_{1}$-norm of quantum coherence. The Schmidt rank analysis confirmed that, unlike the GHZ state where the loss of a single particle severs all connections (collapsing the system to a product state), the Dicke state maintains a bipartite Schmidt rank of $R=2$ for all excitation numbers $0 < k < n$ under single-qubit projective measurement. This mathematically validates the topological persistence of the link. Furthermore, we identified quantum coherence as the physical mechanism underlying this robustness, introducing the concept of link fluidity. We demonstrated that the stability of the topological link is not intrinsic to rigid pairwise bonds, but is rather an emergent property of the delocalized superposition. The derived coherence, $C_{l_{1}} = \binom{n}{k} - 1$, serves as a measure of the system's capacity to redistribute correlations. We found that this fluidity is maximized when the excitation density is balanced ($k \approx \frac{n}{2}$), corresponding to the most densely woven topological structure. Finally, we showed that the measurement dynamics of Dicke states exhibit a recursive self-similarity. The \textit{surgical removal } of a qubit transforms an $n-$ qubit Dicke state into a superposition of $(n-1)-$ qubit Dicke states, preserving the fundamental linking pattern. This suggests that the Dicke state does not merely possess entanglement, but possesses a holographic structural integrity where the global topological properties are encoded within its subsystems. This framework bridges the gap between abstract topology and experimental quantum optics, offering a new perspective on designing robust quantum networks resilient to particle loss. The topological framework established in this paper opens several avenues for future investigation, particularly regarding the resilience of these structures in realistic, open quantum systems. While this work focused on pure states and projective measurements (particle loss), experimental realizations are subject to environmental noise. Future work will investigate how link fluidity degrades under specific noise channels, such as phase damping and amplitude damping. We aim to determine a critical coherence threshold below which the topological link effectively \textit{melts} or disentangles. Our current analysis utilized strong projective measurements in the computational basis. It remains to be seen how the $n-$ Hopf topology responds to weak measurements or Positive Operator-Valued Measures (POVMs). Exploring whether the recursive preservation of the link holds under partial collapse could have significant implications for measurement-based quantum computation. Although we have established a strong phenomenological analogy between Dicke states and Hopf links, a rigorous mathematical mapping to knot polynomials (such as the Jones polynomial or Kauffman bracket) is a necessary next step. Developing a quantum invariant that can distinguish between different classes of multipartite entanglement based strictly on their topological knot equivalence would be a powerful classification tool. The current model relies on the indistinguishability of qubits. We propose extending this analysis to systems with broken symmetry or higher-dimensional particles (qudits). Investigating whether link fluidity can be engineered in non-symmetric states would provide deeper insights into the minimum resources required to construct robust topological quantum memories.

\section{Conflict of Interest:}
The authors declare that they have no known competing financial interests or personal relationships that could have appeared to influence the work reported in this paper.

\section{Data Availability Statement:} Our manuscript has no associated data. 

\newpage
\bibliographystyle{unsrt}
\addcontentsline{toc}{section}{References}
\bibliography{references}

@article{bengtsson2016brief,
  title={A brief introduction to multipartite entanglement},
  author={Bengtsson, Ingemar and Zyczkowski, Karol},
  journal={arXiv preprint arXiv:1612.07747},
  year={2016}
}

@article{thapliyal1999multipartite,
  title={Multipartite pure-state entanglement},
  author={Thapliyal, Ashish V},
  journal={Physical Review A},
  volume={59},
  number={5},
  pages={3336},
  year={1999},
  publisher={APS}
}

@article{bell1964einstein,
  title={On the einstein podolsky rosen paradox},
  author={Bell, John S},
  journal={Physics Physique Fizika},
  volume={1},
  number={3},
  pages={195},
  year={1964},
  publisher={APS}
}

@article{dur2000three,
  title={Three qubits can be entangled in two inequivalent ways},
  author={D{\"u}r, Wolfgang and Vidal, Guifre and Cirac, J Ignacio},
  journal={Physical Review A},
  volume={62},
  number={6},
  pages={062314},
  year={2000},
  publisher={APS}
}

@article{bhattacharyya2025symmetric,
  title={Symmetric and asymmetric tripartite states under the lens of entanglement splitting and topological linking},
  author={Bhattacharyya, Sougata and Roy, Sovik},
  journal={arXiv preprint arXiv:2509.05972},
  year={2025}
}

@inproceedings{bartschi2019deterministic,
  title={Deterministic preparation of Dicke states},
  author={B{\"a}rtschi, Andreas and Eidenbenz, Stephan},
  booktitle={International Symposium on Fundamentals of Computation Theory},
  pages={126--139},
  year={2019},
  organization={Springer}
}

@article{mukherjee2020preparing,
  title={Preparing Dicke states on a quantum computer},
  author={Mukherjee, Chandra Sekhar and Maitra, Subhamoy and Gaurav, Vineet and Roy, Dibyendu},
  journal={IEEE Transactions on Quantum Engineering},
  volume={1},
  pages={1--17},
  year={2020},
  publisher={IEEE}
}

@article{ainley2024multipartite,
  title={Multipartite entanglement for multi-node quantum networks},
  author={Ainley, EM and Agrawal, A and Main, D and Drmota, P and Nadlinger, DP and Nichol, BC and Srinivas, R and Araneda, G},
  journal={arXiv preprint arXiv:2408.00149},
  year={2024}
}

@book{aczel2002,
  title={Entanglement: The Greatest Mystery in Physics},
  author={Aczel, Amir D.},
  year={2002},
  publisher={Four Walls Eight Windows}
}

@book{nielsen2010quantum,
  title={Quantum computation and quantum information},
  author={Nielsen, Michael A and Chuang, Isaac L},
  year={2010},
  publisher={Cambridge university press}
}

@article{horodecki2009,
  title={Quantum entanglement},
  author={Horodecki, R. and Horodecki, P. and Horodecki, M. and Horodecki, K.},
  journal={Rev. Mod. Phys.},
  volume={81},
  number={2},
  pages={865--942},
  year={2009},
  publisher={APS}
}

@book{adams1994,
  title={The Knot Book: An Elementary Introduction to the Mathematical Theory of Knots},
  author={Adams, Colin C.},
  year={1994},
  publisher={American Mathematical Society}
}

@article{dur2000,
  title={Three qubits can be entangled in two inequivalent ways},
  author={D{\"u}r, Wolfgang and Vidal, Guifre and Cirac, J. Ignacio},
  journal={Phys. Rev. A},
  volume={62},
  number={6},
  pages={062314},
  year={2000},
  publisher={APS}
}

@incollection{aravind1997,
  title={Borromean entanglement of the {GHZ} state},
  author={Aravind, P. K.},
  booktitle={Potentiality, Entanglement and Passion-at-a-Distance},
  editor={Cohen, R. S. and Horne, M. and Stachel, J.},
  series={Boston Studies in the Philosophy of Science},
  volume={194},
  pages={53--59},
  year={1997},
  publisher={Springer, Dordrecht}
}

@incollection{greenberger1990,
  title={Going beyond {B}ell's theorem},
  author={Greenberger, Daniel M. and Horne, Michael A. and Zeilinger, Anton},
  booktitle={Bell's Theorem, Quantum Theory and Conceptions of the Universe},
  editor={Kafatos, M.},
  pages={69--72},
  year={1990},
  publisher={Kluwer Academic}
}

@book{peres1995,
  title={Quantum Theory: Concepts and Methods},
  author={Peres, Asher},
  year={1995},
  publisher={Kluwer Academic Publishers}
}

@book{rolfsen1976,
  title={Knots and Links},
  author={Rolfsen, Dale},
  year={1976},
  publisher={Publish or Perish Press}
}

@article{kauffman2016,
  title={Quantum entanglement and topological entanglement},
  author={Kauffman, Louis H. and Lomonaco, Samuel J.},
  journal={New J. Phys.},
  volume={18},
  number={3},
  pages={033032},
  year={2016},
  publisher={IOP Publishing}
}

@article{li2008,
  title={Entanglement and topological entanglement in multiqubit systems},
  author={Li, M. and Fei, S. M. and Wang, Z. X.},
  journal={Phys. Rev. A},
  volume={78},
  number={2},
  pages={022332},
  year={2008},
  publisher={APS}
}

@article{plenio2007,
  title={An introduction to entanglement measures},
  author={Plenio, M. B. and Virmani, S.},
  journal={Quant. Inf. Comput.},
  volume={7},
  number={1},
  pages={1--51},
  year={2007}
}

@article{amico2008,
  title={Entanglement in many-body systems},
  author={Amico, L. and Fazio, R. and Osterloh, A. and Vedral, V.},
  journal={Rev. Mod. Phys.},
  volume={80},
  number={2},
  pages={517--576},
  year={2008},
  publisher={APS}
}

@article{cao2020fragility,
  title={Fragility of quantum correlations and coherence in a multipartite photonic system},
  author={Cao, Huan and Radhakrishnan, Chandrashekar and Su, Ming and Ali, Md Manirul and Zhang, Chao and Huang, Yun-Feng and Byrnes, Tim and Li, Chuan-Feng and Guo, Guang-Can},
  journal={Physical Review A},
  volume={102},
  number={1},
  pages={012403},
  year={2020},
  publisher={APS}
}

@article{roy2023exploring,
author = {Roy, Sovik and Bhattacharjee, Anushree and Radhakrishnan, Chandrashekar and Ali, Md. Manirul and Ghosh, Biplab},
year = {2023},
month = {02},
pages = {},
title = {Exploring quantum properties of bipartite mixed states under coherent and incoherent basis},
volume = {21},
journal = {International Journal of Quantum Information},
doi = {10.1142/S0219749923500107}
}

@article{agrawal2006perfect,
  title={Perfect teleportation and superdense coding with W states},
  author={Agrawal, Pankaj and Pati, Arun},
  journal={Physical Review A—Atomic, Molecular, and Optical Physics},
  volume={74},
  number={6},
  pages={062320},
  year={2006},
  publisher={APS}
}

@article{roy2025environment,
    author = "Roy, Sovik and Kalaiselvan, Aahaman and Radhakrishnan, Chandrashekar and Ali, Md Manirul",
    title = "{Environment Engineering to Protect Quantum Coherence in Tripartite Systems Under Dephasing Noise}",
    eprint = "2412.15082",
    archivePrefix = "arXiv",
    primaryClass = "quant-ph",
    doi = "10.1007/s10773-025-05995-7",
    journal = "Int. J. Theor. Phys.",
    volume = "64",
    number = "5",
    pages = "132",
    year = "2025"
}

@article{dicke1954coherence,
  title={Coherence in spontaneous radiation processes},
  author={Dicke, Robert H.},
  journal={Physical Review},
  volume={93},
  number={1},
  pages={99},
  year={1954},
  publisher={APS}
}

@article{baumgratz2014quantifying,
  title={Quantifying coherence},
  author={Baumgratz, Tillmann and Cramer, Marcus and Plenio, Martin B.},
  journal={Physical Review Letters},
  volume={113},
  number={14},
  pages={140401},
  year={2014},
  publisher={APS}
}

@article{kiesel2007experimental,
  title={Experimental observation of a generalized Dicke state},
  author={Kiesel, Nikolai and Schmid, Christian and T{\'o}th, G{\'e}za and Solano, Enrique and Weinfurter, Harald},
  journal={Physical Review Letters},
  volume={98},
  number={6},
  pages={063604},
  year={2007},
  publisher={APS}
}

@article{wieczorek2009experimental,
  title={Experimental entanglement of a six-photon symmetric Dicke state},
  author={Wieczorek, Witlef and Krischek, Roland and Kiesel, Nikolai and Michelberger, Patrick and T{\'o}th, G{\'e}za and Weinfurter, Harald},
  journal={Physical Review Letters},
  volume={103},
  number={2},
  pages={020504},
  year={2009},
  publisher={APS}
}

@article{toth2007detection,
  title={Detection of multipartite entanglement in the vicinity of symmetric Dicke states},
  author={T{\'o}th, G{\'e}za},
  journal={Journal of the Optical Society of America B},
  volume={24},
  number={2},
  pages={275--282},
  year={2007},
  publisher={Optica Publishing Group}
}

@article{radhakrishnan2024entanglement,
author = {Radhakrishnan, Chandrashekar and Roy, Sovik and Chinnarasu, Ravikumar and Ali, Md. Manirul},
year = {2024},
month = {05},
pages = {},
title = {Entanglement preservation in tripartite quantum systems under dephasing dynamics},
volume = {146},
journal = {Europhysics Letters},
doi = {10.1209/0295-5075/ad3eac}
}

@article{quinta2019cut,
  title={Cut-resistant links and multipartite entanglement resistant to particle loss},
  author={Quinta, Gon{\c{c}}alo M and Andr{\'e}, Rui and Burchardt, Adam and {\.Z}yczkowski, Karol},
  journal={Physical Review A},
  volume={100},
  number={6},
  pages={062329},
  year={2019},
  publisher={APS}
}

\appendix
\newpage

\section{Derivation of the $l_1$-Norm of Coherence for Pure States}
\label{app:coherence_derivation}

In this appendix, we derive the simplified analytical formula for the $l_1$-norm of quantum coherence restricted to pure states \cite{roy2023exploring}. This derivation justifies the calculation $C_{l_1} = (\sum |c_i|)^2 - 1$ used in Section 3.

\subsection{General Definition}
The $l_1$-norm of coherence for a general density matrix $\rho$ in a fixed reference basis $\{|i\rangle\}$ is defined as the sum of the absolute values of the off-diagonal elements:
\begin{equation}
    C_{l_1}(\rho) = \sum_{i \neq j} |\rho_{ij}|
\end{equation}
where $\rho_{ij} = \langle i | \rho | j \rangle$.

\subsection{Pure State Expansion}
Consider a general pure state $|\psi\rangle$ expanded in the reference basis of dimension $d$:
\begin{equation}
    |\psi\rangle = \sum_{i=1}^{d} c_i |i\rangle
\end{equation}
where $c_i$ are complex coefficients satisfying the normalization condition $\sum_{i} |c_i|^2 = 1$. The density matrix $\rho$ for this pure state is:
\begin{equation}
    \rho = |\psi\rangle\langle\psi| = \left( \sum_{i} c_i |i\rangle \right) \left( \sum_{j} c_j^* \langle j| \right) = \sum_{i,j} c_i c_j^* |i\rangle\langle j|
\end{equation}
The matrix elements are given by $\rho_{ij} = c_i c_j^*$.

\subsection{Derivation}
Substituting the matrix elements into the definition of coherence:
\begin{equation}
    C_{l_1}(|\psi\rangle) = \sum_{i \neq j} |c_i c_j^*| = \sum_{i \neq j} |c_i| |c_j|
\end{equation}
We observe that the square of the sum of the absolute values of the coefficients can be expanded as:
\begin{equation}
    \left( \sum_{i} |c_i| \right)^2 = \left( \sum_{i} |c_i| \right) \left( \sum_{j} |c_j| \right) = \sum_{i,j} |c_i| |c_j|
\end{equation}
This double summation over all indices $i,j$ can be split into diagonal terms ($i=j$) and off-diagonal terms ($i \neq j$):
\begin{equation}
    \sum_{i,j} |c_i| |c_j| = \underbrace{\sum_{i} |c_i|^2}_{\text{Diagonal}} + \underbrace{\sum_{i \neq j} |c_i| |c_j|}_{\text{Off-Diagonal}}
\end{equation}
Using the normalization condition $\sum |c_i|^2 = 1$, we can rewrite the equation as:
\begin{equation}
    \left( \sum_{i} |c_i| \right)^2 = 1 + \sum_{i \neq j} |c_i| |c_j|
\end{equation}
Identifying the last term as the coherence $C_{l_1}(|\psi\rangle)$, we arrive at the final formula:
\begin{equation}
    C_{l_1}(|\psi\rangle) = \left( \sum_{i} |c_i| \right)^2 - 1
\end{equation}
This formula allows for the direct calculation of coherence from the state vector coefficients without explicitly constructing the full density matrix.

\newpage
\section{Projective Measurement on the Dicke State $|D_n^{(k)}\rangle$}

This appendix provides the detailed mathematical calculations for the projective measurements performed on the general Dicke state. Due to permutation symmetry, we designate the measured qubit as qubit $1$ without loss of generality.

\subsection{Algebraic Form of the State}
We consider a general Dicke state on $n$ qubits with $k$ excitations:
\begin{equation}
    |D_n^{(k)}\rangle = \frac{1}{\sqrt{\binom{n}{k}}} \sum_{j} P_j \left( |0\rangle^{\otimes (n-k)} |1\rangle^{\otimes k} \right)
\end{equation}
where the sum is over all distinct permutations $P_j$.

\subsection{Measurement on Qubit 1}
We perform a projective measurement in the computational basis $\{|0\rangle, |1\rangle\}$.

\subsubsection{Case 1: Measurement Yields $|0\rangle$}
The projector is $P_0 = |0\rangle\langle 0|_1 \otimes I^{\otimes (n-1)}$.\\\\
\textbf{Application of Projector:}
We apply $P_0$ to the superposition as
\begin{equation}
    P_0 |D_n^{(k)}\rangle = \frac{1}{\sqrt{\binom{n}{k}}} P_0 \left( \sum_{\text{perm}} | \dots \rangle \right)
\end{equation}
We classify the terms in the summation into two types:
\begin{itemize}
    \item \textbf{Type A:} Terms where Qubit 1 is $|0\rangle$. In these terms, all $k$ excitations are distributed among the remaining $n-1$ qubits.
    \item \textbf{Type B:} Terms where Qubit 1 is $|1\rangle$. In these terms, one excitation is fixed on Qubit 1, and the remaining $k-1$ excitations are distributed among the remaining qubits.
\end{itemize}
The action of the projector is:
\begin{equation}
    P_0 |0\rangle\langle 0| (\text{Type A}) = \text{Type A}
\end{equation}
\begin{equation}
    P_0 |0\rangle\langle 0| (\text{Type B}) = 0 \quad (\text{Annihilated})
\end{equation}
Therefore, only Type A terms survive. Each surviving term has the form $|0\rangle_1 \otimes |\phi_j\rangle_{2\dots n}$, where $|\phi_j\rangle$ represents a specific arrangement of $k$ excitations in $n-1$ qubits. The number of such terms is $\binom{n-1}{k}$.\\

Thus:
\begin{equation}
    P_0 |D_n^{(k)}\rangle = \frac{1}{\sqrt{\binom{n}{k}}} |0\rangle_1 \otimes \left( \sum \text{all arrangements of } k \text{ 1's in } n-1 \text{ qubits} \right)
\end{equation}

Using the definition of the Dicke state for $n-1$ qubits ,i.e. $|D_{n-1}^{(k)}\rangle = \frac{1}{\sqrt{\binom{n-1}{k}}} \sum \text{arrangements}$, we get: 

\begin{equation}
   P_0 |D_n^{(k)}\rangle = \frac{\sqrt{\binom{n-1}{k}}}{\sqrt{\binom{n}{k}}} |0\rangle_1 \otimes |D_{n-1}^{(k)}\rangle
\end{equation}

\textbf{Probability:}
The probability is the squared norm:
\begin{equation}
    p(0) = || P_0 |D_n^{(k)}\rangle ||^2 = \frac{\binom{n-1}{k}}{\binom{n}{k}}
\end{equation}
Using the identity $\binom{n}{k} = \frac{n}{n-k}\binom{n-1}{k}$, we can write:
\begin{equation}
p(0) = \frac{\binom{n-1}{k}}{\frac{n}{n-k}\binom{n-1}{k}} = \frac{n-k}{n}
\end{equation}

\textbf{Post-Measurement State:}
Renormalizing the state:
\begin{equation}
   |\psi^{(0)}\rangle = \frac{P_0 |D_n^{(k)}\rangle}{\sqrt{p(0)}} = |0\rangle_1 \otimes |D_{n-1}^{(k)}\rangle
\end{equation}
Tracing out Qubit 1, the remaining system is in the state $|D_{n-1}^{(k)}\rangle$.

\subsubsection{Case 2: Measurement Yields $|1\rangle$}
The projector is $P_1 = |1\rangle\langle 1|_1 \otimes I^{\otimes (n-1)}$.\\\\
\textbf{Application of Projector:}
We apply $P_0$ to the superposition:
\begin{equation}
    P_1 |D_n^{(k)}\rangle = \frac{1}{\sqrt{\binom{n}{k}}} P_1 \left( \sum_{\text{perm}} | \dots \rangle \right)
\end{equation}
We classify the terms in the summation into two types:

\begin{itemize}
    \item \textbf{Type A ( qubit $1$ is $|0\rangle$):} Annihilated ($P_1 |0\rangle = 0$).
    \item \textbf{Type B ( qubit $1$ is $|1\rangle$):} Survives.
\end{itemize}
In the surviving Type B terms, qubit $1$ is fixed at $|1\rangle$. The remaining $n-1$ qubits contain exactly $k-1$ excitations. The number of such terms is $\binom{n-1}{k-1}$.
Thus:
\begin{equation}
    P_1 |D_n^{(k)}\rangle = \frac{1}{\sqrt{\binom{n}{k}}} |1\rangle_1 \otimes \left( \sum \text{all arrangements of } k-1 \text{ 1's in } n-1 \text{ qubits} \right)
\end{equation}
Using the definition $|D_{n-1}^{(k-1)}\rangle = \frac{1}{\sqrt{\binom{n-1}{k-1}}} \sum \text{arrangements}$, we get:
\begin{equation}
    P_1 |D_n^{(k)}\rangle = \frac{\sqrt{\binom{n-1}{k-1}}}{\sqrt{\binom{n}{k}}} |1\rangle_1 \otimes |D_{n-1}^{(k-1)}\rangle
\end{equation}

\textbf{Probability:}
\begin{equation}
    p(1) = \frac{\binom{n-1}{k-1}}{\binom{n}{k}}
\end{equation}
Using the identity $\binom{n}{k} = \frac{n}{k}\binom{n-1}{k-1}$, we simplify:
\begin{equation}
    p(1) = \frac{\binom{n-1}{k-1}}{\frac{n}{k}\binom{n-1}{k-1}} = \frac{k}{n}
\end{equation}

\textbf{Post-Measurement State:}
\begin{equation}
    |\psi^{(1)}\rangle = \frac{P_1 |D_n^{(k)}\rangle}{\sqrt{p(1)}} = |1\rangle_1 \otimes |D_{n-1}^{(k-1)}\rangle
\end{equation}

Tracing out qubit $1$, the remaining system is in the state $|D_{n-1}^{(k-1)}\rangle$.

\subsection{Comparison Between Both the Cases}

Together with the previous result for $P_0$ measurement:
\begin{itemize}
    \item For $P_0$: $p(0) = \frac{n-k}{n}$, post-measurement state $\ket{0}_1 \otimes \ket{D_{n-1}^{(k)}}$.
    \item For $P_1$: $p(1) = \frac{k}{n}$, post-measurement state $\ket{1}_1 \otimes \ket{D_{n-1}^{(k-1)}}$.
\end{itemize}

These satisfy the expected normalization $p(0) + p(1) = 1$ and reflect the symmetric nature of Dicke states.

\newpage
\section{Schmidt Rank Analysis of the Dicke State $|D_n^{(k)}\rangle$}
We want to analyze the entanglement between one specific qubit (Qubit 1) and the remaining $n-1$ qubits in the state $|D_n^{(k)}\rangle$.

\textbf{Case 1: $|D_n^{(0)}\rangle$ (All zeroes)}
\begin{equation}
|D_n^{(0)}\rangle \equiv |0\rangle^{\otimes n}
\end{equation}
This is a product state. \textit{Schmidt Rank = 1}.\\

\textbf{Case 2: $|D_n^{(n)}\rangle$ (All ones)}
\begin{equation}
|D_n^{(n)}\rangle \equiv |1\rangle^{\otimes n}
\end{equation}
This is also a product state. \textit{Schmidt Rank = 1}.\\

\textbf{Case 3: $|D_n^{(k)}\rangle$ where $0 < k < n$}
Using the measurement probabilities calculated above, we can write the state as a superposition of the two outcomes:
\begin{itemize}
    \item If Qubit 1 is $|1\rangle$, the remaining $n-1$ qubits are in $|D_{n-1}^{(k-1)}\rangle$.
    \item If Qubit 1 is $|0\rangle$, the remaining $n-1$ qubits are in $|D_{n-1}^{(k)}\rangle$.
\end{itemize}
Thus, we can express the state as:
\begin{equation}
|D_n^{(k)}\rangle = \sqrt{\frac{n-k}{n}} |0\rangle_1 \otimes |D_{n-1}^{(k)}\rangle + \sqrt{\frac{k}{n}} |1\rangle_1 \otimes |D_{n-1}^{(k-1)}\rangle
\end{equation}

Now, consider the inner product of the two environment states:
\begin{equation}
\langle D_{n-1}^{(k)} | D_{n-1}^{(k-1)} \rangle = \sum_{\text{config}_k} \sum_{\text{config}_{k-1}} \langle \text{config}_k | \text{config}_{k-1} \rangle
\end{equation}
where $|\text{config}_x\rangle$ represents a computational basis state of $n-1$ qubits with exactly $x$ excitations ($1$'s).

\subsection{Key Observation}
Consider any specific computational basis state:
\begin{itemize}
    \item If it appears in $|D_{n-1}^{(k)}\rangle$, it has exactly $k$ ones.
    \item If it appears in $|D_{n-1}^{(k-1)}\rangle$, it has exactly $k-1$ ones.
\end{itemize}
Since these states have different numbers of excitations (i.e., different Hamming weights), they are orthogonal. No computational basis state can appear in both superpositions.
Therefore, every term in the inner product is zero:
\begin{equation}
\langle \text{config}_k | \text{config}_{k-1} \rangle = 0
\end{equation}
Consequently:
\begin{equation}
\langle D_{n-1}^{(k)} | D_{n-1}^{(k-1)} \rangle = 0
\end{equation}

Since the two terms in the decomposition involve orthogonal states for the remaining system, we have a valid Schmidt decomposition with exactly two non-zero coefficients (given $0 < k < n$).
\begin{equation}
\textit{Schmidt Rank} = 2
\end{equation}

The Schmidt coefficients are:
\begin{equation}
\lambda_1 = \sqrt{\frac{n-k}{n}}, \quad \lambda_2 = \sqrt{\frac{k}{n}}\\
\end{equation}

\subsection{Condition for Maximal Entanglement}
For maximal entanglement, the Schmidt coefficients must be equal:
\begin{equation}
\sqrt{\frac{n-k}{n}} = \sqrt{\frac{k}{n}} \implies \frac{n-k}{n} = \frac{k}{n} \implies n-k = k \implies k = \frac{n}{2}
\end{equation}
Therefore, $|D_n^{(k)}\rangle$ is maximally entangled between 1 qubit and the rest if and only if 
\begin{equation}
k = n/2
\end{equation} 
(i.e., exactly half of the qubits are excited).

\newpage
\section{Quantum Coherence Analysis of the Dicke State $|D_n^{(k)}\rangle$}

The $l_1$-norm of quantum coherence is defined as the sum of the absolute values of the off-diagonal elements of the density matrix $\rho$. For a pure state $|\psi\rangle = \sum c_i |i\rangle$, it is given by:
\begin{equation}
    C_{l_1}(|\psi\rangle) = \left( \sum_i |c_i| \right)^2 - 1
\end{equation}

\subsection{Initial State Analysis}
\textbf{State Definition:}
\begin{equation}
    |D_n^{(k)}\rangle = \frac{1}{\sqrt{\binom{n}{k}}} \sum_{j} P_j (|1\rangle^{\otimes k} |0\rangle^{\otimes (n-k)})
\end{equation}
\textbf{Initial Coherence Calculation:}
\begin{itemize}
    \item Number of basis terms: $N = \binom{n}{k}$.
    \item Coefficient for each term: $c = \frac{1}{\sqrt{\binom{n}{k}}}$.
    \item Sum of absolute coefficients:
    \begin{equation}
        \sum |c_i| = \binom{n}{k} \cdot \frac{1}{\sqrt{\binom{n}{k}}} = \sqrt{\binom{n}{k}}
    \end{equation}
\end{itemize}
\textbf{Result:}
\begin{equation}
    C_{l_1}(\text{Initial}) = \left( \sqrt{\binom{n}{k}} \right)^2 - 1 = \binom{n}{k} - 1
\end{equation}

\subsection{Measurement Dynamics and Residual Coherence}
When a projective measurement is performed on Qubit 1 in the computational basis $\{|0\rangle, |1\rangle\}$, the coherence evolves as follows:

\textbf{Case A: Outcome $|0\rangle$}
\begin{itemize}
    \item \textit{Post-Measurement State:} The measured qubit collapses to $|0\rangle$, and the remaining system is in the state $|\phi^{(0)}\rangle = |D_{n-1}^{(k)}\rangle$.
    \item \textit{Residual Coherence:} The system size drops ($n \to n-1$) while the number of excitations remains constant ($k \to k$).
    \item \textit{Calculation:}
    \begin{equation}
        C_{l_1}^{(0)} = \binom{n-1}{k} - 1
    \end{equation}
    \item \textit{Implication:} The coherence decreases ($\binom{n-1}{k} < \binom{n}{k}$), but remains non-zero provided $k < n-1$. Topologically, the link persists.
\end{itemize}

\textbf{Case B: Outcome $|1\rangle$}
\begin{itemize}
    \item \textit{Post-Measurement State:} The measured qubit collapses to $|1\rangle$, and the remaining system transitions to $|\phi^{(1)}\rangle = |D_{n-1}^{(k-1)}\rangle$.
    \item \textit{Residual Coherence:} Both the system size ($n \to n-1$) and the number of excitations ($k \to k-1$) decrease.
    \item \textit{Calculation:}
    \begin{equation}
        C_{l_1}^{(1)} = \binom{n-1}{k-1} - 1
    \end{equation}
    \item \textit{Implication:} The coherence decreases, but remains non-zero provided $k > 1$. The topological link is recursively preserved.
\end{itemize}

\end{document}